\documentstyle[aps,preprint,epsf,epsfig,rotate,prl]{revtex}
\newcommand{\be}{\begin{equation}}
\newcommand{\ee}{\end{equation}}
\newcommand{\ba}{\begin{eqnarray}}
\newcommand{\ea}{\end{eqnarray}}

\draft        
\begin{document}             
\title{ 2D superconductivity with strong spin-orbit interaction }
\author{S. K. Yip}
\address{Institute of Physics, Academia Sinica, Nankang, Taipei 11529, Taiwan}
\maketitle

\begin{abstract}
{\small
We consider superconductivity confined at a two-dimensional
interface with a strong surface spin-orbit (Rashba) interaction.
Some peculiar properties of this system are
investigated.  In particular, we show that an in-plane
Zeeman field can induce a supercurrent flow.

\noindent PACS numbers: 74.20.-z, 74.25.-q, 74.20.Rp, 73.20.-r}

\end{abstract}
\date{\today}
\vspace*{0.2 cm}

  Most superconductors have their underlying crystal structures
and the normal states obeying inversion symmetry. 
This symmetry allows the classification of superconductors 
\cite{Volovik84,Blount85,Yip93} into singlet and triplet
pairing, and correspondingly even and odd symmetry of the
order parameter under sign change of momentum
$ \vec p \to - \vec p$, {\it i.e.} the opposite
sides of the Fermi surface.
 This classification  has played an important role
in our current understanding of superconductors
and their properties.   
 Most "conventional" superconductors
such as Nb and Pb are singlet s-wave \cite{deGennes},
oxide superconductors are likely to be singlet
d-wave \cite{Tsuei00}, whereas superfluid $^3$He is 
triplet p-wave \cite{Leggett75}. 

When inversion symmetry is absent in the normal
state, such classification is no longer possible.  The superconducting
pairing
can thus be neither singlet nor triplet \cite{Gorkov01},
and the order parameter neither even nor odd
under $\vec p \to - \vec p$.  The superconductor can
therefore have rather peculiar physical properties when compared with
those where 
the above mentioned classification can be made.  
This absence of inversion symmetry may be relevant to
some known superconductors.
 (see also references cited in \cite{Edelstein95})
An examination of the list of superconductors
in Table 6.1 of \cite{Poole} shows that, 
e.g., Mo$_3$Al$_2$C (symmetry $P4_132$),
La$_5$B$_2$C$_6$ (symmetry $P4$) and
Mo$_3$P (symmetry $I{\bar 4}$)
are all without inversion centers.
Furthermore,
two-dimensional (2d) surface superconductivities
have been induced by gate electric potentials in C$_{60}$
and some molecular crystals in the field-effect-transistor
geometry \cite{Schon00,Schon01}.
There is no inversion symmetry in these cases
since "up" and "down" are different due to
the electric gates, substrates etc.



Some properties of superconductors without inversion
centers have already
been studied theoretically before (see\cite{Gorkov01,Edelstein95}
and references therein).  For definiteness and
motivated by the last mentioned examples above, we here consider,  
as in \cite{Gorkov01,Edelstein95},
 a 2d superconductor at an interface
with no "up-down" symmetry.  As pointed out there,
one potentially important effect due to the lack of
inversion symmetry in such a geometry is the existence
of a surface spin-orbit coupling or Rashba \cite{Rashba60} term 
in the Hamiltonian
of the form $ - \alpha \hat n \times \vec p \cdot \vec \sigma$.
Here $\hat n$ is the surface normal and $\vec \sigma$
are the Pauli spin matrices.  This term acts like an effective
magnetic field along $\hat n \times \vec p$
and thus splits the spin degeneracy of the electrons at a given
momentum $\vec p$.  The energy difference near the Fermi level
can be large:  in some systems it is known to be of order
$0.1 {\rm eV}$ \cite{LaShell96}, and is therefore expected to be
much larger than the superconducting gap $\Delta$ even for a transition
temperature $ \sim 100 K$.  Rashba
splitting of this magnitude hence is expected to have dramatic effects
on the superconducting properties in these systems.
Some physical consequences due to this spin-orbit
coupling term have been considered in \cite{Gorkov01,Edelstein95}
using Green's function approach.
Gor'kov and Rashba \cite{Gorkov01} calculated the
spin susceptibility in this system.  Edelstein \cite{Edelstein95}
pointed out an interesting magneto-electric effect,
that a spin-polarization can be induced by a 
supercurrent flow.   Here we shall reconsider these
physical properties under the most probable case where
\be
 { p_F^2 \over 2m} >> \alpha p_F >> |\Delta| 
\label{ineq}
\ee 
using simple physical arguments. [ Here $p_F$ is the Fermi momentum
and $m$ is the effective mass.  The definition of $p_F$ will
be made more precise below].  In addition, we give
a more complete description of the magneto-electric
effect in this system.  More precisely, we shall
show the existence of an inverse effect, i.e., 
a supercurrent can be induced by an applied Zeeman field.
The relation of this effect to that proposed by
Edelstein and the possibility of its 
experimental observation is discussed.

We shall then consider a two-dimensional electronic system
lying in the x-y plane.
The one-body part of the Hamiltonian is given by
\be
H^{(1)} = { p^2 \over 2m} - 
         \alpha \hat n \times \vec p \cdot \vec \sigma 
\label{H1}
\ee
with $\hat n = \hat z$.
We shall first summarize some consequences of eq (\ref{H1})
which we shall need below.
As mentioned, the effect of the Rashba term is like a
Zeeman field along $\hat n \times \vec p$.  The   
eigenstates of this spin-dependent part of
the Hamiltonian thus correspond to states with spins
along and opposite to this direction.  
We shall label these spin states by $ |\vec p, +>$ and 
$|\vec p, ->$
respectively.  The spinors for these states can be chosen to be
( by rotating those for an up and down spin by $ -{\pi \over 2} $ along
$\hat p$),
\be
{1 \over {\sqrt 2}} 
\left( 
\begin{array}{c}
1 \\
i e^{i \phi_{\vec p}}
\end{array}
\right)
\qquad {\rm and} \qquad
{1 \over {\sqrt 2}} 
\left( 
\begin{array}{c}
i e^{- i \phi_{\vec p}}\\
1
\end{array}
\right)
\label{spinors}
\ee
where $\phi_{\vec p}$ is the angle between $\hat p$ and the 
$\hat x$ axis in the plane.
The energy of these states at a given momentum $\vec p$
 are given by
$\epsilon_{p,\pm} = { p^2 \over 2m} \mp \alpha |p|$.
For chemical potential $\bar \mu$, the $|+>$ and $|->$ bands
are filled up to Fermi momenta
$p_{F,\pm} = [ ( 2 m \bar \mu ) + m^2 \alpha^2 ] ^{1/2} \pm m \alpha $.
The velocities of the particles are $ { d \epsilon_{p,\pm} \over d p} 
= { p \over m} \mp \alpha $ and different for the $|\vec p, +>$ and
$|\vec p, ->$ particles.  However, at their respective Fermi momenta
the Fermi velocities $v_{F,+}$ and $v_{F,-}$ are equal
and given by $ [ { 2 \bar \mu \over m} + \alpha^2 ] ^{1/2} $. 
The density of states at $\bar \mu$ for the bands are 
$ {\sc N}_{\pm}(0) = { 1 \over 2 \pi \hbar^2}
 { p \over (d \epsilon_{p,\pm}/dp) } 
= { m \over 2 \pi \hbar^2} \left[ 1 \pm { \alpha \over 
[  (2 \bar \mu / m ) + \alpha^2 ] ^{1/2} } \right] $.
They differ slightly (under condition (\ref{ineq}))
by a relative amount of order $\alpha \over \bar \mu$.
In the absence of spin-orbit coupling ($\alpha=0$)
they are both given by ${\sc N}^{0}(0) = { m \over 2 \pi \hbar^2}$.

Let us first consider the spin susceptibility of this
system in the normal state.  For comparison,
we note that the spin susceptibility $\chi^0$ in the absence
of spin-orbit interaction is isotropic and given by 
${ m \over \pi \hbar^2} \mu^2$,
here $\mu$ is the magnetic moment.
This result can be obtained by elementary considerations,
which however we shall summarize since we shall use this
type of argument repeatedly below.  Under a magnetic field B,
the energy of spins aligned (anti-aligned) with the field
is lowered (increased) by $\mu B$.  Since the density
of states is ${m \over 2 \pi \hbar^2}$, the number of particles
(per unit area) for these two species are changed by 
$ \pm { m \over 2 \pi \hbar^2} \mu B$ respectively, giving
a total magnetic moment of $ { m \over \pi \hbar^2} \mu^2 B$ and
hence the Pauli susceptibility given above.

Now we return to the case with $\alpha \ne 0$.
Consider first a magnetic field B perpendicular to the plane
(along $\hat z$).  Since the spins are originally in the plane,
there are no Zeeman energy and thus population changes
for either species.
The Pauli part of the spin susceptibility $\chi^P_{\perp}$ 
therefore vanishes.  However, there is also a Van Vleck contribution
$\chi^{V}_{\perp}$.  Under the $\hat z$ Zeeman field, the 
$|+>$ state is modified to become
\be
|+>' = |+> + { |-><-|\sigma_z|+> \mu B \over 2 \alpha p}
\label{pert}
\ee
according to perturbation theory. 
The expectation value of the $\hat z$ magnetic moment
is given by $'<+|\sigma_z|+>'= {\mu^2 B \over \alpha p}$
(using the spinors in (\ref{spinors})).
 Similar expressions
apply to $|->$.  A net magnetic moment is present
at momentum $\vec p$ if $|+>$ is occupied whereas
$|->$ is not.
The total magnetic moment of the system is therefore given
by
\ba
{\sc M}^{V}_z  &=& { 1 \over 2 \pi \hbar^2} \int_{p_{F-}}^{p_{F+}}
      dp \ p \ { \mu^2 B \over \alpha p}   \nonumber \\
   &=& { \mu^2 \over 2 \pi \hbar^2} { p_{F+} - p_{F-} \over \alpha } B \ .
\ea
Using the expressions for $p_{F \pm}$, we obtain
$\chi^V_{\perp} = { m \over \pi \hbar^2} \mu^2 = \chi^0 $,
the same spin susceptibility
in the absence of spin-orbit coupling. 

\begin{figure}[h]
\epsfig{figure=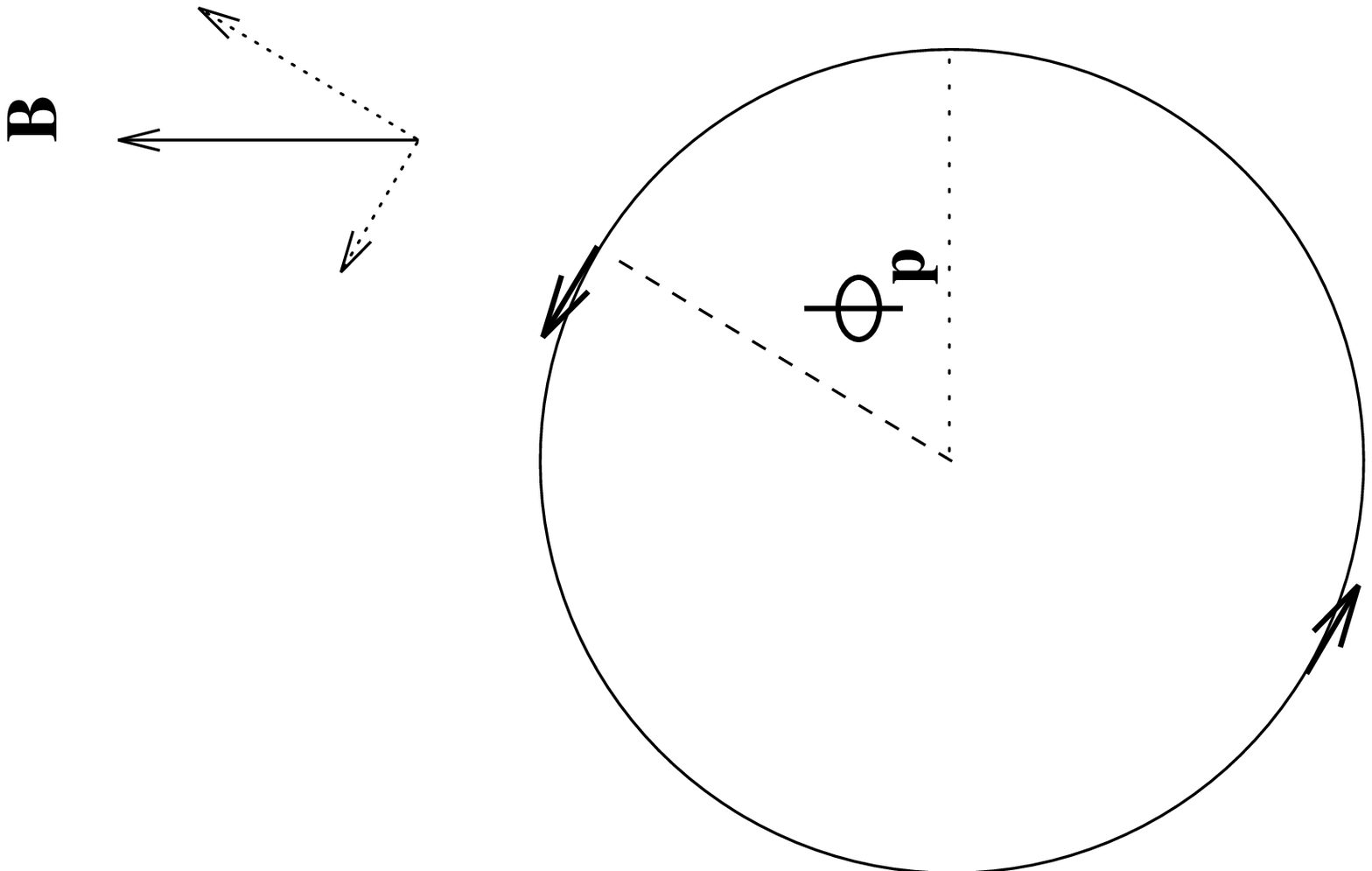,width=3in,angle=-90}
\vskip 0.4 cm
\begin{minipage}{0.45\textwidth}
\caption[]{(1) Spin directions (thick arrows) on the $|+>$
fermi surface at two representative (equal and opposite)
momenta.  These two electrons form a pair in the superconducting
state. (2) An applied magnetic field ${\vec B}$ is resolved into
components parallel and perpendicular to the spin direction.
The $|->$ spins are not shown in this figure.}

\label{fig:spin}
\end{minipage}
\end{figure}

Now consider a magnetic field in the plane, e.g., along
the $\hat y$ axis.  To calculate the spin susceptibility
it is convenient, for each momentum $\hat p$, to resolve
$\vec B$ into components parallel and perpendicular
to the momentum direction $\vec p$ (see Fig. \ref{fig:spin}).
  The former (latter)
field is perpendicular (parallel) to the original spin
direction, and can only give rise to a Van Vleck
(Pauli) contribution to the net magnetic moment.
One easily finds, using arguments as in the
last two paragraphs, the results
\be
\chi^{P}_{\parallel}  = 
   ( {\sc N}_{+}(0) + {\sc N}_{-}(0) ) \mu^2 / 2
  = \chi^0 / 2
\label{chillP}
\ee
and
\be
\chi^{V}_{\parallel} = \chi^0 /2 \  \ .
\label{chillV}
\ee
The $1/2$ in eq (\ref{chillP}) and
(\ref{chillV}) are due to angular averages.
We obtain finally $\chi_{\parallel} =
\chi^{V}_{\parallel} + \chi^{P}_{\parallel} = \chi^0$.
Hence the spin susceptibility is not affected at all
by the Rashba term.  This result has been obtained
also in \cite{Gorkov01}.

Now we consider the superconducting state.  We shall 
consider the case where the Cooper pairing occurs
between the $\pm \vec p$ particles from the same band,
{\it i.e.}, between $ | \vec p, +> $ and $| - \vec p, +>$
on the one hand (see Fig. \ref{fig:spin})
and  between $ | \vec p, -> $ and $| - \vec p, ->$
on the other.  We shall also limit ourselves to
the case where the energy gaps $\Delta_{\pm}$ may be different for
the two bands but isotropic in momentum space.
That the pairing occurs only within the same
band is reasonable since we assume that
the energies associated with the pairing $\Delta_{\pm}$
are much less than the energy separation between the
two bands $2 \alpha p_{F \pm}$ for a given momentum
$\vec p$ near $p_{F \pm}$ (see eq (\ref{ineq})).
The assumption of this pairing is consistent with that in \cite{Gorkov01}.
  We shall not
justify it here and shall simply consider its physical
consequences.  Situations where $\Delta_{\pm}$ are $\hat p$
dependent seem also possible and the following results
can be generalized to these cases by simple arguments.

Consider now the spin susceptibility in the superconducting 
state, first for a magnetic field perpendicular to the plane.
In this case argument as in the normal state shows that
the Pauli susceptibility vanishes.  The Van Vleck susceptibility,
being generated by virtual processes to states
with energy separations much larger than $\Delta_{\pm}$
 (if (\ref{ineq}) applies), is little affected.  
We get therefore
$\chi^V_{\perp}(T) =  \chi^V_{\perp}(T > T_c) = \chi^0$
and thus $\chi_{\perp} = \chi^0$ independent of the superconducting
transition.

Now consider a magnetic field in the plane.  For the contribution
from the pair $\pm \vec p$,   we argue as in
the normal state and resolve the magnetic field into
components parallel and perpendicular to $\hat p$. 
The former again gives only a 
Van Vleck contribution unaffected by the superconducting transition, thus
the total Van Vleck susceptibility 
$\chi^V_{\parallel}(T) 
= \chi^0 /2 $
as in the normal state.
The field component perpendicular to $\hat p$ 
again gives only a Pauli contribution,
which can be evaluated by arguments as in the case of superfluid $^3$He
\cite{Leggett75}.  Consider first the $|+>$ band. 
In the absence of the magnetic field
the Hilbert space for
$\pm \vec p$ consists of four possible states: 
 ground pair (GP) with energy $0$,
(two) broken pair state (BP) with energy $E_{p+}$
 $ = \sqrt { \xi_{p+}^2 + |\Delta_{+}|^2 }$
[ here $\xi_{p+} \equiv \epsilon_{p+} - \bar \mu$ is
the normal state quasiparticle energy relative to the chemical potential ]
corresponding to occupied (empty) $ |\vec p,+>$ and empty
(occupied) $|-\vec p,+>$, and
excited pair (EP) with energy $2 E_{p+}$.
Under the magnetic field, these energies are modified
to become $0$, $E_{p}-h_p$, $E_{p}+h_p$, $2 E_{p}$
where $h_p = \mu B {\rm cos} \phi_{\vec p}$,
since the magnetic moment of $|\vec p, +>$ along the field
 is $\mu {\rm cos} \phi_{\vec p}$. 
(We are leaving out the $+$ subscripts for the moment for
easier writing.)
The net magnetization
along the field direction is therefore
\be
(\mu {\rm cos} \phi_{\vec p} ) \
[ e^{-(E_p-h_p)/T} - e^{-(E_p+h_p)/T} ] \  / \  Z 
\label{thav}
\ee
where $Z \equiv 1 + e^{-(E_p-h_p)/T} + e^{-(E_p+h_p)/T} + e^{-2E_p/T} $
is the partition function.
For small magnetic field, this reduces to $\mu^2 B {\rm cos}^2 \phi_{\vec p}
\ { 1 \over 4 T} {\rm sech}^2 { E_p \over 2 T} $.
The total magnetization of the $|+>$ band is given by
summing over $\vec p$, which is the same
as multiplying by $ { 1 \over 2} {\sc N}_{+} (0)$, 
integrate over $\xi_p$ and average over $\phi_{\vec p}$.
 (the $1/2$ factor is to avoid
counting the same pair twice).  
The angular average gives a factor of $1/2$.  
We obtain the contribution to the Pauli susceptibility
$ \mu^2 {\sc N}_{+}(0) Y(T,\Delta_{+})/ 4$ from this band.
Here $Y(T,\Delta) \equiv
\int d \xi { 1 \over 4 T} {\rm sech}^2 { E_p \over 2 T} $
is the Yosida function.
The total Pauli susceptibility from both bands is thus
$ \chi^P_{\parallel}(T) = \mu^2 [ {\sc N}_{+}(0) Y(T,\Delta_{+}) + 
     ( + \leftrightarrow - ) ] / 4 $.
The full susceptibility is given by
$\chi_{\parallel}(T) = \chi^V_{\parallel} + \chi^P_{\parallel}(T)$.
  If 
$\Delta_{+} = \Delta_{-}$, 
we get $\chi_{\parallel}(T) = \chi^0 ( 1 + Y(T,\Delta))/2$.
The above results agree with those in \cite{Gorkov01}
under the condition (\ref{ineq}). 

Now we turn to the electro-magnetic effects.  We shall show
that an applied Zeeman field in the plane, say along $\hat y$,
can produce a supercurrent flow along $\hat x$ in
the superconducting state.  To demonstrate this
we shall first consider the normal state and show that
the net current vanishes due to the cancellation of two contributions
which can be identified as "Pauli" and "Van Vleck".
These two contributions are due respectively to the
change in occupation and the quantum-mechanical wave-function of
the particles
as in the case of spin susceptibility.  We shall then show
that the cancellation no longer holds in the superconducting
state, giving rise to the mentioned net supercurrent.

Consider now a magnetic field $B$ along $\hat y$ in the normal state.
The physical situation is as shown in Fig \ref{fig:spin}.
Let us first consider the Pauli contribution from the  $|+>$ band.  The
$\hat y$ magnetic moment of the electron at $\vec p$ is given by
$\mu {\rm cos} \phi_{\vec p}$.  Hence the extra number of 
occupied states (per unit area and per unit angle)
due to the magnetic field
 with momentum near $\hat p$ is given
by ${\sc N}_{+}(0) ( \mu B {\rm cos} (\phi_{\vec p}) )$. 
These electrons have velocity $v_{F+}$ along $\hat p$.  
Hence the current along $\hat x$ is equal to
the angular average of 
${\sc N}_{+}(0) (v_{F+} \mu  B) {\rm cos}^2 (\phi_{\vec p})$
{\it i.e.}, ${ 1 \over 4 \pi \hbar^2} p_{F+} ( \mu B) $, using 
$ {\sc N}_{+}(0) v_{F+} = { p_{F+} \over 2 \pi \hbar^2 } $.
[ This Pauli contribution is therefore
due to the fact that states with $p_x > 0$ are more
likely to be occupied than $p_x < 0$ under the field $B_y$.]
The reverse situation applies for the $|->$ band.
The total (number) current density 
from both bands due to these population changes 
is given by 
\be
J^P_x = { 1 \over 4 \pi \hbar^2} (p_{F+} - p_{F-} ) \mu B_y
\label{JP}
\ee
The superscript $P$ denotes that this is the "Pauli" contribution.
In addition to this, there is also a "Van Vleck" contribution.
The velocity of an electron at $\vec p$, given by
$\vec v = {\partial \epsilon \over \partial \vec p}$,
is actually $ {\vec p \over m} \hat 1 + \alpha (\vec n \times \vec \sigma)$
and thus an operator in spin space.  In particular
$v_x = { p_x \over m} - \alpha \sigma_y$.  Under the magnetic field
$B_y$, the $|+>$ state is modified as in 
eq (\ref{pert}) with $\sigma_z \to \sigma_y$.
  Hence the expectation value of $v_x$ is
given by $ ( { p \over m}  - \alpha ) {\rm cos} (\phi_{\vec p})
       - { \mu B \over p } |<-|\sigma_y|+>|^2$.
The first term is the velocity of the $|+>$ particle
in the absence of $B$ and its contribution to the
current was taken into account by the Pauli term evaluated
before.  The second term, equals 
$ - {\mu B \over p} {\rm sin}^2  (\phi_{\vec p})$, is present
due to the modification of the state under the Zeeman field.
We shall call its contribution to the current a Van Vleck contribution 
analogous to the case for the spin susceptibility.
A net Van Vleck contribution at $\vec p$ is present only if $|+>$ is
occupied whereas $|->$ is empty.  The total Van Vleck
current is thus
$J_x^V =  { 1 \over 2} { 1 \over 2 \pi \hbar^2} \int_{p_{F-}}^{p_{F+}} 
  dp \ p \  \left( - {\mu B \over p} \right) $
where the factor $1/2$ arises from angular average.
We hence obtain 
\be
J_x^V = -   { 1 \over 4 \pi \hbar^2} (p_{F+} - p_{F-} ) \mu B_y
\ee
giving $J_x = J^P_x + J^V_x = 0$ in the normal state as claimed.
[ It can be easily shown that $J^P_y$ and $J^V_y$ both vanish
 due to angular average over the fermi surface].
The vanishing of the total current is reasonable since
otherwise dissipation is expected in the presence of disorder.

In the superconducting state the calculation of $J$ is 
similar to that of the susceptibility.  The Van Vleck
contribution $J^V$ is unaffected, while
the Pauli contribution has to be multiplied by
the Yosida functions.  We therefore get
\be
J_x(T) = - \kappa B_y
\label{Jx}
\ee
where
\be
\kappa(T) =  { \mu \over 4 \pi \hbar^2} \ \left[ 
    p_{F+} ( 1 - Y(T,\Delta_+) ) - p_{F-} ( 1 - Y(T,\Delta_-)) \right]
\label{kappa}
\ee

We can similarly investigate the effect
pointed out by Edelstein \cite{Edelstein95}, {\it i.e.},
the generation of a magnetic moment by a phase gradient.
Under a phase gradient $\nabla \Phi$, say along $\hat x$, 
the Cooper pairing is no longer between $\pm \vec p$
but rather between $ \vec p + \vec q /2 $ and 
$ - \vec p + \vec q/2$, where $\vec q = \hbar ( \nabla \Phi)$.
Let us first calculate the net magnetic moment at $T=0$.
In this case the magnetic moment is the same as that of
a Fermi sphere (circle) shifted in momentum space by $\vec q /2$.
The total moment can be found by summing over all the excess
( over $\vec q = 0$) moments over the fermi surface(s).  For the 
$|+>$ particles, the number of extra particles along $\hat p$ is given by
${\sc N}_{+}(0) [ \epsilon (\vec p + \vec q/2) - \epsilon (\vec p)]
 = {\sc N}_{+}(0) v_{F+} q {\rm cos} (\phi_{\vec p}) / 2$
since the quantity between the square bracket is the difference
in energy between the particles on the new and old fermi surfaces.
These particles carry a $\hat y$ magnetic moment of 
$\mu {\rm cos} (\phi_{\vec p})$
per particle.  Hence the total $\hat y$ magnetic moment 
from the $|+>$ band is given by
$ \mu {\sc N}_{+}(0) v_{F+} q / 4$.  Therefore the total contribution from
the two bands is 
\be
{M}_y(T=0) = { \mu \over 8 \pi \hbar^2} ( p_{F+} - p_{F-} ) q 
\label{M0}
\ee
It can be easily seen that the $\hat x$ magnetic moment vanishes
due to angular average over $\hat p$.

The above result eq (\ref{M0}) is when all electrons remained paired.
At finite temperatures, we need to take into account the 
contribution from broken pairs.  For this it is essential to
note that, under the phase gradient, the energies
for a broken pair with particles occupied at $\vec p$ is given
by $E_{\vec p} + \vec v_F(\vec p) \cdot \vec q /2$, where
$E_{\vec p}$ is the energy given before for no phase gradient.
The thermal-averaged magnetic moment for the $\pm \vec p$ states is given
by an expression similar to (\ref{thav}) in the susceptibility calculation
 with $ -h_p \to \vec v_F(\vec p) \cdot \vec q /2 
=  v_{F+} q {\rm cos} (\phi_{\vec p}) / 2 $, 
giving the final result
$- { \mu \over 8 \pi \hbar^2} ( p_{F+} q) Y(T,\Delta_{+})$.
[ This negative contribution from the quasiparticles is therefore physically
due to the "backflow", that it is easier to thermally
excite quasiparticles with momentum opposite to the superfluid flow.
These particles have a net magnetic moment along $-\hat y$ for
the $|+>$ band.] 
A similar expression applies for the $|->$ band. 
Combining these with eq (\ref{M0}),  we therefore
have finally
\ba
{M}_y (T) &=& 
 { \mu \over 8 \pi \hbar^2} \left[ p_{F+} ( 1 - Y(T,\Delta_{+}) )
     - p_{F-} ( 1 - Y(T,\Delta_{-}) \right] q  \nonumber \\
  &=& {\kappa \over 2} q_x 
\label{My}
\ea
with $\kappa(T)$ already defined in eq (\ref{kappa}).
For $T$ near $T_c$, we can perform an expansion in 
$\Delta$.  
[ $ 1 - Y \to { 7 \zeta(3) \over 4 \pi^2} {\Delta^2 \over T_c^2}$]
 Our expression then agrees with that given
by Edelstein \cite{Edelstein95}, who investigated the effect
only near $T_c$.

The two magneto-electric effects above are related.  They
are connected by the fact that there is a cross-term in the
free energy density $F (T; q_x, B_y)$ given by
$- {\kappa (T) \over 2} q_x B_y$.  Eq (\ref{My}) and  (\ref{Jx}) 
can be reproduced by using the relations
$M_y = - \partial F / \partial B_y$ and
$J_x = 2 \partial F / \partial q_x$.

Generally, the current $J_x$ and magnetization ${M}_y$ 
are given by the constitutive equations 
\ba
J_x  &=&  \rho_s  { q_x \over 2 m}  - \kappa B_y 
\label{con1} \\
M_y  &=&  { \kappa \over 2} q_x     + \chi_{\parallel} B_y
\ea
where $\rho_s$ is the superfluid (number) density.

The supercurrent induced by the in plane Zeeman field 
 given in eq (\ref{Jx}) can be
sizeable and should be experimentally observable.
The order of magnitude of
the electric current $I$ at $T << \Delta$ for a sample
of width $w$ induced by the magnetic field is given by 
\be
\left[{ I \over Amp} \right] \  
  = \  10^{-2} \left[ {\alpha p_F \over \bar \mu} \right]
  \ \left[ { B \over G} \right]
  \ \left[ { 1 \over l / \AA} \right]
  \ \left[ { w \over cm} \right]
\ee
where we have defined a length $l$ of order of interparticle
distance through the two dimensional number density $n$ by
$n = l^{-2}$. 
If ${\alpha p_F \over \bar \mu}$ is not too small,
say $\sim 0.1$,   a current of order of $mA$ seems
easily achievable for samples of $mm$ size under 
a magnetic fields of order $100G$ if $l \sim 10 \AA$, say.  
Measurement of this current seems much easier than
the induced magnetization predicted by Edelstein \cite{Edelstein95}. 

I thank John C. C. Chi for a useful correspondence.



\end{document}